\documentclass[%
 aip,
 amsmath,amssymb,
preprint,%
]{revtex4-1}

\usepackage{graphicx}% Include figure files
\usepackage{dcolumn}% Align table columns on decimal point
\usepackage{bm}% bold math
\usepackage[mathlines]{lineno}% Enable numbering of text and display math
\usepackage[mathlines]{lineno}% Enable numbering of text and display math
\usepackage[utf8]{inputenc}
\usepackage[T1]{fontenc}
\usepackage{mathptmx}
\usepackage{xcolor}

\begin{document}

%\preprint{AIP/123-QED}

\title[Direct laser acceleration in varying plasma density profiles]{Direct laser acceleration in varying plasma density profiles}
% Force line breaks with \\

	\author{R. Babjak}
	\email[]{robert.babjak@tecnico.ulisboa.pt}
	%\thanks{}
	\affiliation{GoLP/Instituto de Plasmas e Fusão Nuclear, Instituto Superior Técnico, Universidade de Lisboa, Lisbon, 1049-001, Portugal}
	\affiliation{Institute of Plasma Physics, Czech Academy of Sciences, Za Slovankou 1782/3, 182 00 Praha 8, Czechia}

 	\author{B. Martinez}
	\affiliation{GoLP/Instituto de Plasmas e Fusão Nuclear, Instituto Superior Técnico, Universidade de Lisboa, Lisbon, 1049-001, Portugal}

    \author{M. Krus}
	%\thanks{}
	\affiliation{Institute of Plasma Physics, Czech Academy of Sciences, Za Slovankou 1782/3, 182 00 Praha 8, Czechia}

	\author{M. Vranic}
	\affiliation{GoLP/Instituto de Plasmas e Fusão Nuclear, Instituto Superior Técnico, Universidade de Lisboa, Lisbon, 1049-001, Portugal}

\date{\today}% It is always \today, today,

\begin{abstract}
Direct laser acceleration has proven to be an efficient source of high-charge electron bunches and high brilliance X-rays. However, an analytical description of the acceleration in the interaction with varying plasma density targets is still missing. Here, we provide an analytical estimate of the maximum energies that electrons can achieve in such a case. We demonstrate that the maximum energy depends on the local electron properties at the moment when the electron fulfills the resonant condition at the beginning of the acceleration. This knowledge enables density shaping for various purposes. One application is to decrease the required acceleration distance which has important implications for multi-petawatt laser experiments, where strong laser depletion could play a crucial role. Another use for density tailoring is to achieve acceleration beyond the radiation reaction limit. We derive the energy scaling law that is valid for arbitrary density profile that varies slowly compared with the betatron period. Our results can be applied to electron heating in exponential preplasma of thin foils, ablating plasma plumes, or gas jets with long-scale ramp-up.  
\end{abstract}

\maketitle

\section{Introduction}

The progress in ultra-short laser pulse technology has allowed a significant development in the acceleration of particles based on the interaction of such pulses with plasmas. The density of gaseous plasma is low enough to allow the laser pulse to propagate through it and accelerate particles by various mechanisms. The most established mechanism of the electron acceleration is the laser wakefield acceleration \cite{faure2004,geddes2004,mangles2004,leemans2014} (LWFA) which is dominant for the relativistically intense laser pulses 10s of fs short. The propagation of such a short laser pulse creates a longitudinal plasma wave that under favorable conditions allows for acceleration of mono-energetic electron beams in order of several GeV and total charge 10s of pC. The highest energy obtained experimentally up-to-date is 7.8 GeV \cite{gonsalves2019}. 

\begin{figure}[t]
	\includegraphics[width=0.99\textwidth]{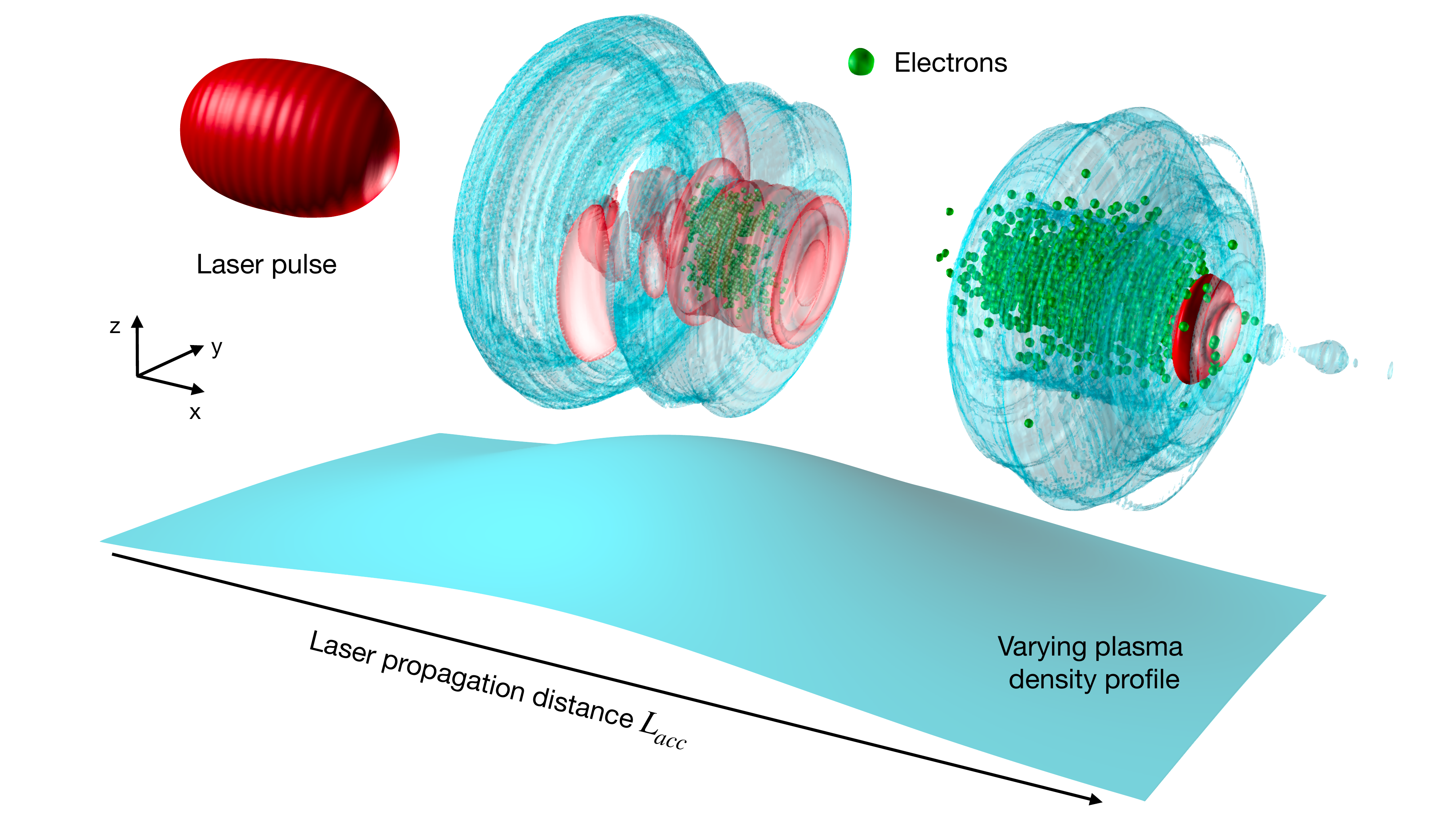}
	\caption{Interaction setup: a relativistic Gaussian linearly polarized laser pulse propagates through a varying density underdense plasma. The direct laser acceleration of electrons occurs in the self-created plasma channel. Within the ion channel, tens of percent of the laser energy can be transferred through the interaction into multi-GeV electrons and collimated X-ray radiation. }
\label{fig:setup_fig} 
\end{figure}

Longer laser pulses with hundreds of fs duration propagating through underdense or near-critical plasmas are favorable for electron acceleration mechanism called direct laser acceleration (DLA) \cite{pukhov1998,pukhov1999,babjak2024}. The ponderomotive force expels electrons from the high-intensity region creating an ion channel attracting electrons towards the channel axis at the center of the laser pulse \cite{pukhov1999,arefiev2012,khudik2016,huang2017,valenta2024}. The interaction setup is shown in Fig. \ref{fig:setup_fig}. Electrons injected inside the ion channel perform betatron oscillations at the same time as they oscillate in the field of the laser pulse and the resonance between the two types of oscillations can lead to the acceleration of electrons to energies exceeding the vacuum limit. The energy spectrum of electrons accelerated by the DLA is broadband and the total accelerated charge exceeds the charge delivered by the LWFA by orders of magnitude \cite{hussein2021,babjak2024}. Electrons can be also partially accelerated by the surface plasmons created on the channel walls which can potentially serve as an injector for the DLA \cite{raynaud2007,naseri2012,marini2021,shen2021}. Electrons exceeding vacuum energy limit due to the effect of DLA have also been observed experimentally \cite{gahn1999,mangles2005,kneip2008,willingale2013,rosmej2019,rosmej2020,willingale2018,hussein2021,rinderknecht2021}. In recent experiments, 100s of nC electron bunches were observed after the acceleration by the DLA in low-density plasmas \cite{hussein2021,shaw2021}. The simultaneous contribution of the DLA and LWFA to the electron energy is also possible and has been studied both theoretically and experimentally \cite{shaw2017,shaw2018,king2021,lamac2021,miller2023}. The combination of both mechanisms could potentially lead to the acceleration of electrons to energies exceeding 10 GeV\cite{aniculaesei2023}. This indicates that the DLA has the potential to be a promising source of multi-GeV electrons with total charge in order of 100s of nC using upcoming multi-petawatt laser facilities \cite{zou2015,webber2017,tanaka2020}. 

The DLA electrons are suitable for applications such as x-ray and gamma generation \cite{chen2013,stark2016,jansen2018,rosmej2021,gunther2022}, ion acceleration \cite{snavely2000,wilks2001} or electron-positron pair creation \cite{chen2009,vranic2018,he2021,he2021_2,amaro2021,martinez2022} due to the high number of accelerated electrons. This can be also advantageous for seeding pair cascades \cite{bellkirk2008,nerush2011,grismayer2017} in the future. Even though the direct laser acceleration has been mostly studied for setups when the laser pulse interacts with the constant density targets, it can be the dominant absorption mechanism in situations when the density changes as the laser pulse propagates. 

For example, the DLA is relevant in the scenario when the laser pulse interacts with the expanding corona of the imploded capsule in a fast ignition setup \cite{tabak1994,tabak2005,kemp2014}. As the intense laser pulse propagates through the ablated plasma, its density increases up to the critical density and heats the electrons. The understanding of the process of high-intensity laser absorption will enable better control of suprathermal electrons coming from the interaction in the corona and consequently more efficient ignition of the core. 

Furthermore, it has been shown that the efficiency of ion acceleration through the TNSA mechanism increases with the presence of preplasma \cite{nuter2008,raffestin2021}. It happens due to the enhanced laser absorption into hot electron population via DLA during the laser propagation through the preplasma and via stochastic heating \cite{mendoca1983,sentoku2002,paradkar2011,babjak2021} in the field of the standing wave after the laser reflection from the high-density foil. It has been shown, that if a strongly relativistic laser pulse interacts with the foil, the preplasma is inevitably present even if very good contrast is assumed \cite{holec2018}. This means that the absorption of the laser by exponentially increasing preplasma via DLA in front of the foil should be accounted for in the TNSA regime. 

It has been also demonstrated experimentally that the density gradient can be beneficial for the electron acceleration itself \cite{hussein2021}. Even during the interaction with low-density plasma, the total charge exceeding 100 nC with energies up to 500 MeV was obtained which contradicts the expectations that low-density targets result in low total charge. It has been shown, that long propagation distance enabling the charge loading for a longer time can compensate for the low background charge, resulting in 100s of nC electron bunches\cite{babjak2024}. Also, the ablation of high-Z foils and consequent interaction with the plasma plume of varying resulted in very efficient electron injection \cite{cohen2024}.

Recently, we have found that the direct laser acceleration at constant density profiles can be optimized in terms of electron cut-off energy by matching the laser spot size to the transverse oscillation amplitude of the most energetic electrons \cite{babjak2024}. The experimental campaign at OMEGA EP laser indicates, that such a non-trivial trade-off between the laser intensity and spot size strongly influences the acceleration, in agreement with our results \cite{tang2024}. Applying the optimization strategy to the regime of multi-PW lasers will enable the creation of multi-GeV electron bunches with a total charge of over-ponderomotive electrons exceeding 100 nC. Even though the direct laser acceleration at varying densities is relevant for many laser-plasma applications, the theoretical description for the acceleration is missing. This limits the development of mentioned applications and the interpretation of experimental results relies solely on simulations. In this paper, we present the analytical description of the interaction of an electron accelerated by a plane wave in an ion channel. We show that the maximum energy that electrons can achieve is a general conserved quantity defined by the initial conditions at the beginning of the acceleration. We also derive the generalized energy scaling law for arbitrary density profile, which varies slowly with respect to betatron oscillations. For example, the scaling allows for predicting the cut-off energy of the electron distribution resulting from the interaction of a laser pulse with an exponentially increasing preplasma of a thin foil. It is also valid in the case of laser propagation through the expanding plasma plume where the plasma density profile has a Gaussian shape \cite{hussein2021}. By understanding the acceleration in varying densities, we can shape the density profile to shorten the acceleration distance required in the multi-PW lasers experiments. For example, we demonstrate using Quasi-3D particle-in-cell (PIC) simulations, that such a target can be used for an acceleration of electrons to energies over 5 GeV by a 5 PW laser in 1 mm of plasma. This strategy might be crucial to overcome the problem of laser depletion. Last but not least, we demonstrate that tailoring the density profile can help overcome the radiation reaction limit on the maximum electron energy. This will propel the electrons to exceed 10 GeV in a configuration with a 10 PW laser, suppressing the radiation losses expected at the final stage of acceleration. 

The paper is organized as follows. In section \ref{sec:constant_theory} we summarize previously known theoretical description of the DLA for the constant density, section \ref{sec:theory} presents our generalization of the description of the acceleration and conserved quantities and in section \ref{sec:scaling} we derive the general scaling law for the cut-off energy for arbitrary density profile. In section \ref{sec:tp_test} we verify the analytical finding by numerically integrating the equations of motion and in section \ref{sec:pic_scaling} we verify the derived energy scaling laws using particle-in-cell simulations. In section \ref{sec:pic_shaping} we demonstrate how density profiles can be shaped to shorten the acceleration distance and in section \ref{sec:pic_rr} we propose a strategy to avoid the effects of radiation reaction using density shaping.  

\section{Theoretical description of the direct laser acceleration} 

\subsection{Constant density}\label{sec:constant_theory}

To describe the acceleration of electrons via DLA for the varying density regime, it is convenient to introduce the known theory for the constant density case. 

The analytical description of the direct laser acceleration assumes a plane wave propagating at the speed of light in the $\boldsymbol{\hat{x}}$ direction within a plasma channel. The field in a plasma channel is assumed to be $\boldsymbol{E_c} =m_e\omega_p^2/2e \boldsymbol{r} $, where $\boldsymbol{r}= y\boldsymbol{\hat{y}}+z\boldsymbol{\hat{z}}$ \cite{khudik2016,arefiev2016,arefiev2016_2,vranic2018b}. Low density of the plasma allows us to neglect the magnetic component of the channel field and assume the luminal phase velocity ($v_{ph}\approx c$) of the laser. We also neglect the effects of a sheath field created at density down-ramps \cite{tang2024}. Under the mentioned assumptions, equations describing the motion of an electron are 

\begin{equation}\label{eq:dpx}
\frac{dp_x}{dt} = -ev_yB_z
\end{equation}
\begin{equation}\label{eq:dpy}
\frac{dp_y}{dt} = -\frac{1}{2}m_e\omega_p^2y+e(v_xB_z-E_y)
\end{equation}
\begin{equation}\label{eq:dgamma}
\frac{d}{dt}(\gamma m_ec^2) = -e (\vec{v}\vec{E}) = -ev_y(\frac{m_e\omega_p^2}{2e}y+E_y),
\end{equation}

where the components of electric and magnetic field correspond to the field of a linearly polarized laser pulse and the terms proportional to $\omega_p^2$ express the contribution of the ion channel fields.

Equations have a conserved quantity (an integral of motion) in a form

\begin{equation}\label{eq:i}
I = \gamma - \frac{p_x}{m_ec} + \frac{\omega_p^2y^2}{4c^2},
\end{equation}

which is a conserved quantity of the system if the plasma density is constant throughout the interaction.

When an electron propagates in the ion channel, it performs betatron oscillations with the frequency $\omega_{\beta}=\omega_p/\sqrt{2\gamma}$ \cite{pukhov2002}. Simultaneously, it undergoes the motion defined by the interaction with the field of a plane wave with the frequency $\omega_L$. Effective acceleration by the DLA mechanism is possible when the betatron resonance condition is fulfilled. The Doppler shifted frequency of the laser field in the luminal case is $\omega_D = \omega_L(1-v_x/c)$.  The resonance becomes possible if the Doppler-shifted laser frequency and the betatron frequency are close ($\omega_D \simeq \omega_{\beta}$) \cite{pukhov1999}. 

Whether an electron gets in the resonance is determined by the resonant amplification condition \cite{arefiev2014,khudik2016}

\begin{equation}\label{eq:res}
\varepsilon_{cr} = \frac{\omega_pa_0}{\omega_0I^{3/2}},
\end{equation}

This limits the possible combinations of $\omega_p$ and $y_0$ that can lead to the resonant interaction, where $\varepsilon_{cr}$ is a dimensionless constant with a value of $0.2$ in the most optimal scenario \cite{babjak2024}. It is not uncommon to observe higher values of $\varepsilon_{cr}$, resulting in non-optimal acceleration and consequently lower than expected electron energies. This can happen due to several reasons including too tight laser focusing, insufficient laser propagation distance to accelerate electrons to the maximum possible energy or short duration of the laser pulse. 

It can be also understood in a way that minimum plasma frequency for electrons to be resonantly accelerated can be expressed as $\omega_p^{\rm{min}}/\omega_0 = \varepsilon_{cr}I^{3/2}/a_0$. If the plasma density is lower than $\omega_p^{\rm{min}}$, resonant acceleration is not possible and the value of $a_0$ needs to be increased to observe efficient acceleration by the DLA.

The maximum energy that can be achieved for an electron accelerated by the DLA with a given initial transverse distance $y_0$ at a plasma channel with a constant plasma frequency $\omega_p/\omega_0$ can be estimated analytically \cite{khudik2016}. We will refer to the quantity as to the $\gamma_{\rm{max}}$ and it can be expressed as 

\begin{equation}\label{eq:gmax}
\gamma_{max} \simeq 2I^2\left(\frac{\omega_0}{\omega_p}\right)^2
\end{equation}

for the electrons located far from the channel axis which have the potential to achieve the highest energies. The maximum energy of electrons close to the channel axis with lower $y_0$ obeys a slightly adjusted scaling described previously in the literature \cite{khudik2016}.

Another aspect of the DLA necessary to consider during the acceleration description is how fast can be the energy from the laser and background field transferred to the electron.  In other words, what propagation distance is needed for an electron in resonance to achieve the energy $\gamma_{max}$ in the case of an optimal resonance.  It has been shown that the energy gain depends on the integral of motion, plasma frequency and laser $a_0$ as\cite{jirka2020,khudik2016}

\begin{equation}\label{eq:dgdt}
\frac{d\gamma}{dt} \simeq \frac{8m_ec^2}{\pi^2} \frac{a_0\omega_p}{\sqrt{I}}.
\end{equation}

Inversely proportional energy gain to the integral of motion means that electrons with the highest $\gamma_{max}$ are the ones with the slowest acceleration which results in a trade-off between the maximum energy achievable and the acceleration distance that is needed for an electron to be accelerated.

\subsection{Maximum energy at the varying density}\label{sec:theory}

In many experimentally relevant scenarios, the plasma density is not constant and previously discussed properties valid for the constant density DLA need to be generalized. Since $\omega_p=\omega_p(x)$ now depends on the position, the integral of motion $I$ expressed by Eq. (\ref{eq:i}) is no longer conserved during the interaction. Another unknown is whether the maximum energy of an electron can be still estimated using Eq. (\ref{eq:gmax}), since both quantities $I$ and $\omega_p$ vary as the laser pulse propagates.

\begin{figure}[h]
		\includegraphics[width=.7\textwidth]{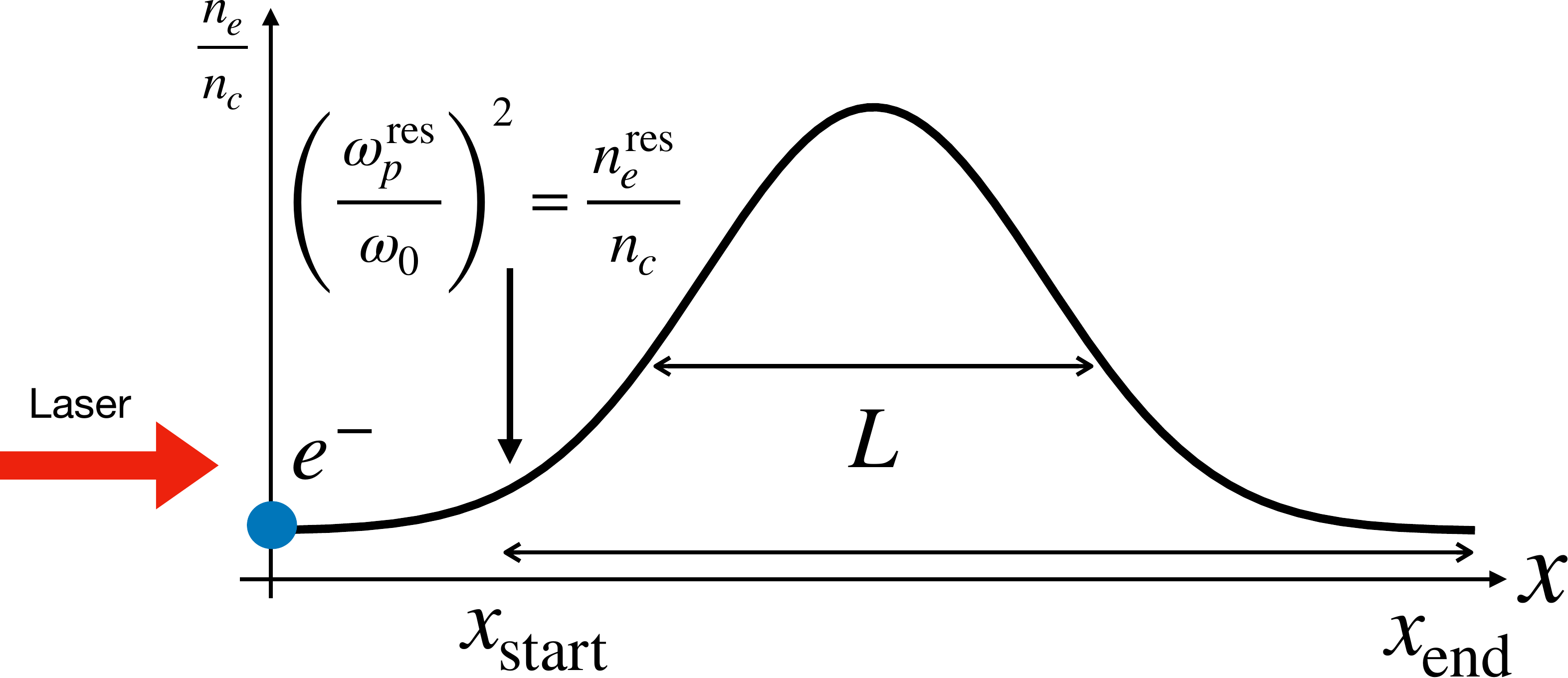}
		\caption{Example of a varying plasma density profile. An electron is initially located at the low-density region and starts to co-propagate with the laser pulse towards higher densities. After reaching a high enough plasma density $n_e^{\rm{res}}$ at position $x_{\rm{start}}$, the particle starts the resonant motion and continues to be resonantly accelerated by the DLA to higher than the ponderomotive limit energies. }
    \label{fig:sketch} 

	\end{figure}

We assume an electron to be initially located at the low-density region. When it starts to interact with the propagating laser pulse, it gets pushed toward the higher densities, see Fig. \ref{fig:sketch}. The motion in low-density plasma is generally non-resonant and electron energies can be described by the vacuum interaction while at the same time electron performs betatron oscillations. When the local density at the position of an electron is sufficiently high to fulfill the resonant condition Eq. (\ref{eq:res}) they start to be accelerated to over-ponderomotive energies by the DLA. The start of a resonant acceleration corresponds to the position $x_{\rm{start}}$ in Fig. \ref{fig:sketch}.

The motion of an electron before the resonance during its co-propagation with the laser pulse in the ion channel can be described as follows. An electron performs betatron oscillations, while it oscillates in the field of the laser pulse simultaneously. The laser pulse is also responsible for the longitudinal velocity and consequent drift toward the higher-density regions. 

Betatron oscillations are described by the harmonic oscillator equation

\begin{equation}\label{eq:lho}
\frac{d^2y}{dx^2} + \frac{\omega_p(x)^2}{2\gamma} y= 0,
\end{equation}

where an electron propagates along the laser propagation direction $\boldsymbol{\hat{x}}$, assuming $p_x \gg p_y$. The density profile $\omega_p(x)^2$ is slowly varying density profile compared to the betatron wavelength.

The equation can be solved analytically for various profiles. For example, linearly increasing density has a solution in the form of the Airy function, with an asymptotic solution for the oscillations amplitude $Y(x)\approx y_0 x^{-1/4}$. For exponentially increasing density profile, the solution can be found in the form of Bessel functions of a zeroth and first kind.  Generally speaking, we can claim that for the varying density profile, the amplitude of betatron oscillations changes asymptotically as $\approx x^{p}$, where the value of exponent $p$ depends on the chosen profile while the amplitude decreases for increasing density and vice versa. 

The first step towards the generalized description is to estimate how the $I$ changes with the varying density profile. We obtain the expression 

\begin{equation}\label{eq:didt}
 \frac{dI}{dt}=\frac{y^2}{4}\frac{d}{dt}\omega_p^2.
\end{equation}

We can see explicitly that if $\omega_p$ is constant, $I$ is conserved, which is consistent with the known description of the constant density scenario. 

To understand the behavior of the local $\gamma_{max}$ in varying density profiles, we can calculate the derivative for each locally assigned value and obtain

\begin{equation} \label{eq:dg_deriv}
\begin{split}
\frac{d\gamma_{max}}{dt} %&= 2\frac{dI}{dt}I\left(\frac{\omega_0}{\omega_p}\right)^2 + %I^2\omega_0^2 \frac{d}{dt}\frac{1}{\omega_p^2}\\
%&=  2I\frac{\omega_0^2}{\omega_p^4} \left[ \frac{y^2\omega_p^2}{4c^2} -\frac{I}{2} %\right] \frac{d}{dt} \omega_p^2\\
&= 2 I \frac{\omega_0^2}{\omega_p^4} \left[ U_p - R \right] \frac{d}{dt} \omega_p^2,
\end{split}
\end{equation} 

 where we used Eq. (\ref{eq:didt}) and expressed $U_p=y^2\omega_p^2/4c^2$ as a potential part of the integral of motion and $R = \gamma-p_x/m_ec$ commonly referred to as a dephasing rate.
 
To further understand the term's meaning, we can use that electron is strongly relativistic during the acceleration $p_x\gg p_y$. This allows us to express the dephasing rate using $p_y \simeq m_ec\sqrt{2\gamma I}\sin(\psi)$ as
\begin{equation}
R = \gamma - \frac{p_x}{m_ec} \simeq \frac{1}{2\gamma} \left(\frac{p_y}{m_ec}\right)^2 = I\sin^2(\psi),
\end{equation}
where $\psi$ is a phase of betatron oscillation. 

The transverse betatron oscillations can be expressed as $y = Y\cos(\psi) = 2\sqrt{I}c\cos(\psi)/\omega_p$. Using this, the value of $U_p$ can be expressed as
\begin{equation}
U_p = \frac{y^2\omega_p^2}{4c^2} = 
%\frac{\omega_p^2}{4c^2} \frac{4Ic^2}{\omega_p^2}\cos^2(\psi)=
 I\cos^2(\psi)
\end{equation}

Substituting previous estimates into Eq. (\ref{eq:dg_deriv}) we obtain

%\begin{equation} \label{eq:dg_deriv2}
%\begin{split}
%\frac{1}{2} \frac{d\gamma_{max}}{dt} &=  I \frac{\omega_0^2}{\omega_p^4} \left[  %I\cos^2(\psi)-  I\sin^2(\psi) \right] \frac{d}{dt} \omega_p^2 \\
%&=  I^2 \frac{\omega_0^2}{\omega_p^4} \cos(2\psi) \frac{d}{dt} \omega_p^2 
%\end{split}
%\end{equation} 
%By this, we get to the final expression for the derivative of $\gamma_{max}$
\begin{equation}\label{eq:dgdt_final}
\frac{d\gamma_{max}}{dt} = 2I^2 \frac{\omega_0^2}{\omega_p^4} \cos(2\psi) \frac{d}{dt} \omega_p^2 
\end{equation}

Note that this expression is valid only after the particle reaches the resonance so the condition $p_x \gg p_y$ is satisfied. 

The most important conclusion to take from Eq. (\ref{eq:dgdt_final}) is that after the electron reaches the resonance, the mean value of $\gamma_{max}$ does not change and is defined by its value at the moment of resonance.  The reason is, that the derivative oscillates around 0 with twice the betatron frequency which means that overall the value of $\gamma_{max}$ does not change in resonance since
\begin{equation}\label{eq:consv}
\left\langle \frac{d\gamma_{max}}{dt}\right\rangle \sim \langle\cos(2\psi)\rangle = 0
\end{equation}

In other words, the maximum energy of an electron being accelerated at varying density profile is determined by the local quantities ($n_e$, $I$) at the moment when it gets in the resonance. Particles can propagate after achieving the resonance through higher-density regions without the change of change of the maximum achievable energy $\gamma_{\rm{max}}$. Since the energy transfer from the laser to electrons is faster at higher densities in the DLA regime according to Eq. (\ref{eq:dgdt}), increasing the plasma density after the electron reaches the resonance will result in electron acceleration at a shorter distance. We demonstrate using PIC simulations in section \ref{sec:pic_shaping} that such tailoring of plasma density can decrease the acceleration distance needed for acceleration.

\subsection{Varying density energy scaling law }\label{sec:scaling}

The estimate for the maximum allowed energy $\gamma_{\rm{max}}$ is not enough for the full description of the interaction, because the interaction length is not always sufficiently long to reach the maximum energy. To estimate the energy of electrons accelerated at a given density profile after a certain distance, one needs to use the formula for the maximum attainable accelerating power (Eq. \ref{eq:dgdt}) and integrate it over the corresponding propagation distance.

We can again assume low plasma density, luminal laser pulse propagation and particles in the resonance with velocities close to $c$, which allows a change of variable $x \approx ct$.  To perform the integration, we need to know how the $I$ changes during the propagation as a function of position $x$, since it is no longer a conserved quantity. However, we have shown that the local value of $\gamma_{\rm{max}}$ oscillates around the constant value defined at the moment of achieving the resonance, see Eq. (\ref{eq:consv}). We can therefore assume $\gamma_{\rm{max}}$ to be constant on distances larger compared to the betatron wavelength. If we express $I$ as

\begin{equation}\label{eq:i_wp}
I(x) = \sqrt{\frac{\gamma_{max}}{2}}\frac{\omega_p(x)}{\omega_0},
\end{equation}

its value depends only on the evolution of plasma frequency $\omega_p$ and the value of $\gamma_{\rm{max}}$ which is constant. 

As mentioned in the previous analysis, the value of $\gamma_{\rm{max}}$ for varying density profiles is defined by the properties of each electron at the moment when it fulfills the resonance condition Eq. (\ref{eq:res}) by the combination of instantaneous plasma frequency at the position $\omega_p$ and the distance from the axis $y_0$. The threshold amplification condition can be used to express the maximum transverse amplitude of an electron capable of achieving the resonance $y_{\rm{max}}$\cite{babjak2024}

\begin{equation}\label{eq:ymax}
    y_{\rm{max}} =\frac{2c}{\omega_p}\sqrt{\left(\frac{a_0\omega_p}{\varepsilon_{cr}\omega_0}\right)^{2/3}-1}.
\end{equation}

	\begin{figure}[h]
		\includegraphics[width=.6\textwidth]{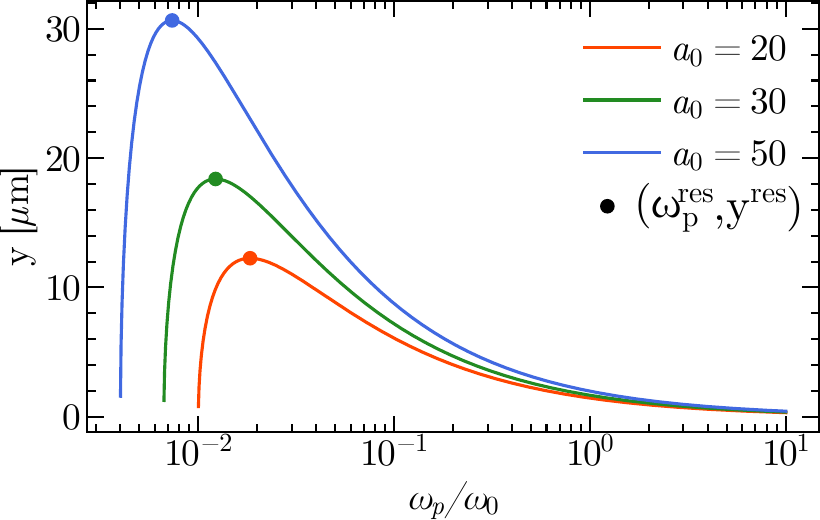}
		\caption{The maximum resonant initial distance from the channel axis of electrons given by the Eq. (\ref{eq:ymax}) for different values of $a_0$. The point represents the combination of plasma frequency $\omega_p^{res}$ (Eq. \ref{eq:wp_res}) and the transverse distance of resonant electrons $y^{res}$ (Eq. \ref{eq:y_res}). Electrons with such initial conditions can reach the maximum possible energy for a laser with given $a_0$ given by the Eq. (\ref{eq:g_res}) in case they succeed to get resonant.}
    \label{fig:ymax} 

	\end{figure}

Electrons initially located at $y_{max}$ achieve the highest energies because $\gamma_{\rm_{max}}\sim y_0^4$. Electrons initially located further from the axis than the distance $y_{max}$ don't fulfill the resonance condition and they are accelerated by the higher-order resonance \cite{arefiev2024} that is less effective and can be neglected \cite{khudik2016}. Therefore, the highest energy electrons can be found by the condition $dy_{\rm{max}}/d\omega_p=0$ which maximizes the distance from the axis of the most energetic electrons, see Fig. \ref{fig:ymax}. This leads to the formula for the resonant plasma frequency of the most energetic electrons expressed as

\begin{equation}\label{eq:wp_res}
    \frac{\omega_p^{\rm{res}}}{\omega_0}= \frac{\varepsilon_{cr}}{a_0}\left(\frac{3}{2}\right)^{3/2}.
\end{equation}

By substituting the resonant plasma frequency into the maximum transverse distance $y_{\rm{max}}$ (Eq. \ref{eq:ymax}) we obtain

\begin{equation}\label{eq:y_res}
    y^{\rm{res}}=\frac{\sqrt{2}c}{\omega_p^{\rm{res}}}.
\end{equation}

Knowing of the maximum transverse distance of resonant electrons $y^{\rm{res}}$ and the plasma frequency $\omega_p^{res}$ at which they achieve the resonance, we can express their $\gamma_{max}$ by substituting Eq.(\ref{eq:wp_res}) and (\ref{eq:y_res}) into the Eq. (\ref{eq:gmax}). This leads to the estimate for the energy of the most energetic electrons that can be directly used in the scaling as

\begin{equation}\label{eq:g_res}
    \gamma_{max}^{\rm{res}} = \left(\frac{4a_0}{3\varepsilon_{cr}}\right)^2.
\end{equation}

The knowledge of the maximum resonant energy $\gamma_{max}$ for a laser with given $a_0$ allows us to express $I$ in Eq. (\ref{eq:dgdt}) using Eq. (\ref{eq:i_wp}) and we obtain

\begin{equation}\label{eq:sl_final}
\frac{d\gamma}{dx} \sim a_0 \left(\frac{2}{\gamma_{max}}\right)^{1/4}\sqrt{\frac{\omega_p(x)}{\omega_0}}.
\end{equation}

This gives us the prescription for the instantaneous energy gain over the element of a propagation distance $dx$. By substituting the arbitrary slowly varying density profile in the formula and integrating it, one obtains the energy of an electron as a function of $\gamma_{max}$ defined by the properties at the moment of achieving the resonance, laser intensity $a_0$ and the propagation distance. The formula can be integrated for any profile where the effects of radiation reaction or superluminal laser phase velocity can be neglected for the regime of underdense plasma assumed in our work. 

As an example, we assume the density profile of an exponential preplasma of a thin foil in a form $n_e/n_c=\exp(x/L)$, where an electron starts interacting with the laser pulse at $x_{\rm{start}}<0$ and finishes the acceleration at $x=0$. This gives us an estimate for the cut-off energy of the hot electron distribution created in the preplasma of the expanded thin foil or for the electron distribution accelerated in the corona of a fast ignition pellet. The maximum energy of an electron interacting with such a profile with the scale length $L$ can be predicted as

\begin{equation}
    E[\rm{MeV}] = 5.08 a_0 L[\rm{\mu m}] \left(\frac{2}{\gamma_{max}}\right)^{1/4} \left[1-\exp\left(\frac{x_{\rm{start}}}{4L}\right)\right],
\end{equation}

where $a_0$ is the peak normalized laser amplitude and the $\gamma_{max}$ can be estimated by the Eq. (\ref{eq:g_res}). If the preplasma does not extend up to the position $x_{\rm{start}}$, the value of $\gamma_{max}$ and integration lower boundary needs to be adjusted accordingly. The value of $\varepsilon_{cr}$ is expected to be equal to 0.2 if the favorable conditions of the optimal acceleration are satisfied \cite{arefiev2014,khudik2016,babjak2024}. 

\section{Test particle verification of the theory}\label{sec:tp_test}

\begin{figure}[h]
		\includegraphics[width=.80\textwidth]{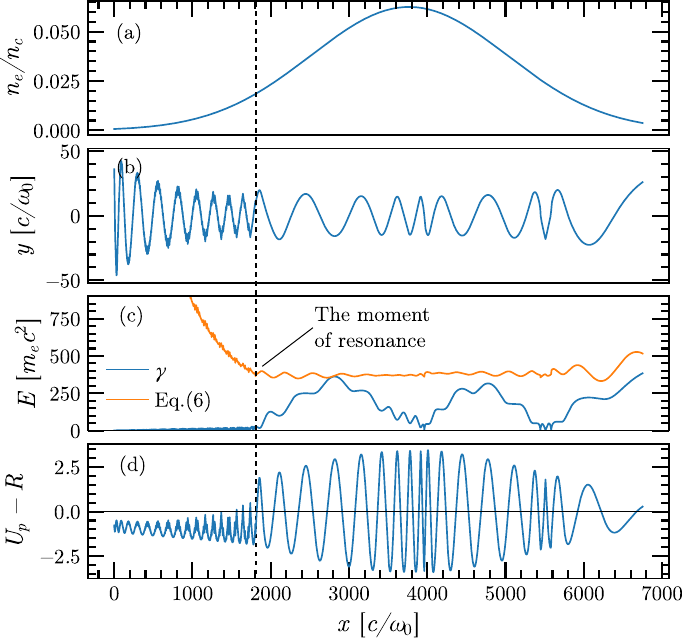}
		\caption{ Evolution of an example test particle trajectory in case of a Gaussian density profile. A laser pulse with a peak field amplitude $a_0=15$ enters the box from the left-hand side and propagates towards the right side. a) Plasma density profile in the test particle simulations. b) Trajectory of an electron propagating through the plasma channel in the field of the laser pulse c) Energy of the electron $\gamma$ at a given position and the value of the $\gamma_{\rm{max}}$ evaluated using local properties from the test particle track from Eq. (\ref{eq:gmax}). After the electron reaches the resonance, the value of $\gamma_{\rm{max}}$ oscillates around the constant value defined at the beginning of the resonant acceleration highlighted by the vertical line. d) The difference between the potential energy of an electron $U_p$ and the dephasing rate $R$ which is proportional to the $d\gamma_{max}/dt$, see Eq. (\ref{eq:dg_deriv}) and (\ref{eq:consv}).}
    \label{fig:tp_gradient}
	\end{figure}

To verify the previously derived conclusions, we performed test-particle (TP) simulations of an electron interacting with a laser field in a plasma channel described by Eq. (\ref{eq:dpx}) - (\ref{eq:dgamma}). The electron is initially placed at a low-density region of a Gaussian density profile as the laser pulse with a Gaussian temporal envelope starts to interact with it. The laser pulse is transversely a plane wave and pushes the test electron forward in the positive $x$ direction. The example interaction scenario is shown in Fig. \ref{fig:tp_gradient} where an electron interacts with the laser pulse with a peak intensity $a_0=15$. 

Fig. \ref{fig:tp_gradient} a) shows the plasma density profile assumed in the test particle simulation. The local plasma density influences the field of an ion channel as $\boldsymbol{E_c} =m_e\omega_p^2/2e \boldsymbol{r} $. An electron is initially placed at $x=0$ and a laser pulse with the Gaussian temporal profile enters the simulation box from the left-hand side.

Fig. \ref{fig:tp_gradient} b) shows the trajectory of an electron, which is pushed in the positive $x$ direction by the laser. In the first part of the trajectory, the electron performs betatron oscillations off-resonance with a decreasing transverse oscillation amplitude (as we discussed previously, within a varying density plasma, the betatron oscillations have a lower amplitude at higher plasma densities). Simultaneously, electrons perform rapid oscillations in the field of a laser pulse. The center position of the particle's rapid oscillations is defined by the betatron phase.  At the position $x \approx 1800~c/\omega_0$, the type of oscillation visibly changes, since the electron gets in the resonance and starts to be accelerated. The frequency of oscillations varies with the evolution of the electron energy which is shown in panel c).  

In panel c), the blue curve corresponds to the energy of the electron from the simulation and the orange one is the theoretical value of $\gamma_{\rm{max}}$ calculated by substituting the values from the test particle trajectory in the Eq. (\ref{eq:gmax}). Before the resonance ($x<1800~c/\omega_0$), the energy of the electron is defined mostly by the field of the laser pulse and follows the vacuum solution of the electron interacting with a plane wave.  After the resonance ($x>1800~c/\omega_0$), the energy gain to energies over the vacuum limit can be observed. The maximum possible energy for the electron is equal to the value of $\gamma_{\rm{max}}$ defined at the moment of the resonance, which is in agreement with our analytical analysis in section \ref{sec:theory}. We note here that $\gamma_{\rm{max}}$ is a constant only when averaged over the resonant cycle. In reality, the attainable values oscillate around this average, according to Eq. (\ref{eq:dgdt_final}). At first, the value of $\gamma_{\rm{max}}$ decreases, since resonance is not reached yet and Eq. (\ref{eq:dgdt_final}) is not valid in this regime. However, when the threshold amplification condition Eq. (\ref{eq:res}) becomes fulfilled, the electron starts to be resonantly accelerated and the mean value of $\gamma_{\rm{max}}$ is conserved. 

We showed that $d\gamma/dt\sim U_p-R$, which is the difference between the part of $I$ associated with the electrostatic potential and the dephasing rate $R$. We remind here that $I$ can be expressed as $I=R+U_p$. If we plot the quantity in Fig.  \ref{fig:tp_gradient} d) using trajectories from TP simulations, we can see that after the resonance is reached, the quantity oscillates around zero, which results in no overall change of the $\gamma_{\rm{max}}$ keeping it constant during the resonance. This is in agreement with Eq. (\ref{eq:dgdt_final}) and consequently Eq. (\ref{eq:consv}). The average value of $U_p-R$ is not zero before the resonance because conditions for the derivation of $d\gamma_{\rm{max}}/dt$ are not fulfilled, as $p_x\gg p_y$ is not valid for small $a_0$ before the resonance is reached. 

\begin{figure}[h]
	\includegraphics[width=.90\textwidth]{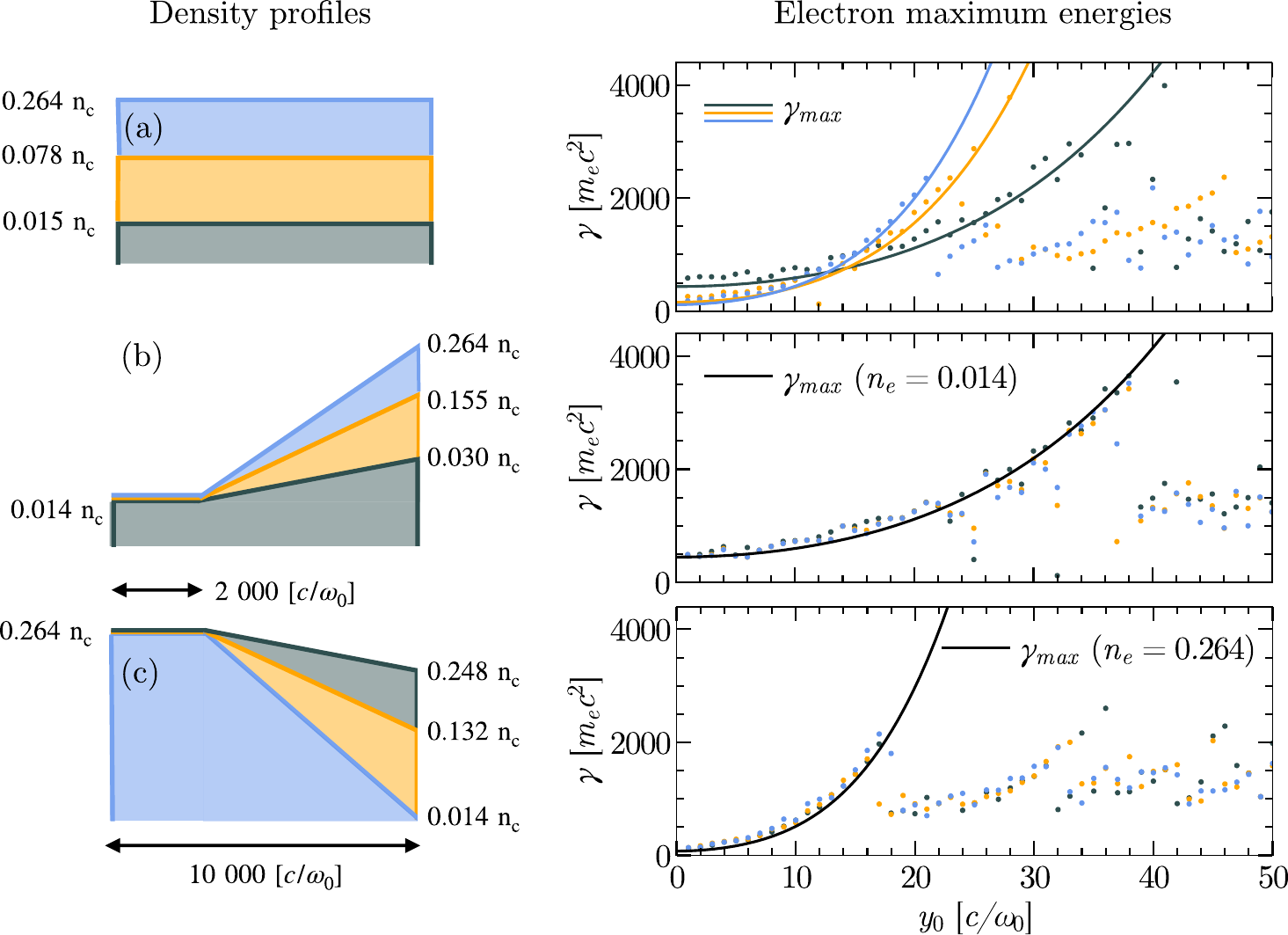}
	\caption{Demonstration of the conserved maximum allowed energy $\gamma_{max}$ after entering resonance for the varying density plasma. We have integrated equations of motion for different density profiles (left column) and compared the maximum energies that electrons obtained. These maxima are a function of an initial distance from the axis $y_0$ (right column), just like for flat density profiles shown in panel (a). The dependence of the maximum energy for constant density profiles is Eq. (22) derived by Khudik et al\cite{khudik2016}. In panels (b) and (c), the constant density part at the beginning is common for all three cases presented and it is long enough for all electrons to reach the resonance in this section of the plasma. Consequently, it defines the maximum achievable energy for all density profiles in (b) and (c), where the analytical prediction is depicted by the black line and all results agree with this prediction. We note that as shown in (a), the different constant density profiles result in different maximum achievable energies. This comparison shows that it is the plasma density at resonance that determines the asymptotic particle energies. }
	\label{fig:tp_gmax}
\end{figure}

The main conclusion of the description presented above is the fact that the density at the moment of achieving resonance defines the maximum achievable energy $\gamma_{\rm{max}}$ for each electron. Since after reaching the resonance the mean value of the $\gamma_{\rm{max}}$ is conserved, the maximum achieved energy is not dependent on the density profiles after that moment. To verify this prediction, we run the test particle simulation scan for three different kinds of profiles, see Fig. \ref{fig:tp_gmax}.

Fig. \ref{fig:tp_gmax} (a) demonstrates the acceleration of electrons at a constant density profile. For each density, the maximum energy of electrons with a given initial transverse distance from the axis $y_0$ varies differently in agreement with the theory\cite{khudik2016}. We can also see that the $y_0$ of the most energetic electrons is higher for lower densities, which agrees with Eq. (\ref{eq:ymax}) and has important implications for laser focusing \cite{babjak2024}.

In Fig. \ref{fig:tp_gmax} (b), all three chosen profiles share the constant density of $0.014~n_c$ for the distance 2000 $c/\omega_0$ (corresponding to 318 $ \rm{\mu m}$ for the laser of 1 micron wavelength) and after, the constant density is followed by the linear increase of the density with varied steepness. Electrons achieved the same energies for all three profiles considered and the energy scaling agrees with the prediction for the plasma density $0.014~n_c$ at the beginning (left-hand side) of the plasma profile. The reason is that all electrons achieved resonance while the plasma density was constant at $0.014~n_c$. This defined their maximum achievable energy, in agreement with Eq. (\ref{eq:consv}) and the discussion about the $\gamma_{\rm{max}}$ conservation. This demonstrates that the shape of the density profile does not influence the maximum achievable energy limit that is set at the beginning of the acceleration.

Plasma profiles with a low density at the entrance followed by a ramp-up can therefore be used for decreasing the required acceleration distance. Low density of the plasma results in high-energy electrons, but the acceleration distance needed to achieve such energy is generally long \cite{babjak2024}. To reduce the acceleration distance (and overcome the possible problems caused by the laser depletion) we propose to start the acceleration at the low density to ensure the highest possible maximum energy $\gamma_{\rm{max}}$, and later increase the density to increase the local acceleration rate defined by the Eq. (\ref{eq:dgdt}). By tailoring the density profile we can thus achieve the highest possible energies at the shortest distances.

The Fig. \ref{fig:tp_gmax} (c) shows the constant density profiles followed by linearly decreased density. The maximum energy reached for those profiles is again the same for all density profiles and agrees with energy scaling for the density at the beginning of the acceleration $0.264~n_c$ where electrons reached the resonance. Such density profiles can be beneficial for the acceleration by lasers with a total power of 10 PW and higher to suppress the effects of radiation reaction and achieve the acceleration at the shortest possible distance to overcome the laser depletion before reaching the maximum energy $\gamma_{\rm{max}}$. This will be discussed in more detail in section \ref{sec:pic_shaping}.

\section{Particle-in-cell simulations}\label{sec:pic}
\subsection{Energy scaling law verification}\label{sec:pic_scaling}

As a laser pulse propagates through the plasma, it undergoes strong focusing and defocusing and it is subject to many instabilities connected with the nonlinearity of the interaction. For that reason, we perform particle-in-cell (PIC) simulations to verify the validity of the cut-off energy scaling law and the maximum energy conservation. 

In the section \ref{sec:scaling} the energy scaling was derived for arbitrary slowly varying density profile. We will compare it with the PIC simulation results where the density profile was chosen as Gaussian. In the simulation the laser pulse with a Gaussian transverse profile with $W_0=10$ microns and dimensionless field amplitude $a_0=10$. The duration of the laser pulse was 600 fs with an envelope defined by the symmetrical polynomial function that rises as $10\tau^3-15\tau^4+6\tau^5$, where $\tau=\sqrt{2}/\tau_0$ and $\tau_0$ is the pulse duration in FWHM. After it was focused, it interacted with the Gaussian density profile $n_e = 0.02\exp[(x-x_0)^2/L^2]$ along the distance $\approx$ 1500 $\mu \rm{m}$ ($x_0=750 \rm{\mu m}$). The scale length of the density profile was $L=433$ $\mu \rm{m}$. The plasma density at the focus was $n_e=9\times 10^{-4} n_c$.  The Eq. (\ref{eq:sl_final}) can be integrated for the Gaussian density profile and after dividing it by the factor of 2 to account for the shape of the laser pulse it results in the scaling law in the form 

\begin{equation}\label{eq:sl_gauss}
E(x)[\rm{MeV}] = 2.25a_0 L[\rm{\mu m}]\left(\frac{2n_e^{max}}{\gamma_{max}n_c}\right)^{1/4}\left[\rm{erf}\left( \frac{x-x_0}{2L}\right) - \rm{erf}\left( \frac{-x_0}{2L}\right)\right],
\end{equation} 

where $n_e^{max}=0.02n_c$ is the highest value of a Gaussian density profile and the dimensionless value of $\gamma_{max}$ is predicted by the Eq. (\ref{eq:g_res}).

\begin{figure}[h]
	\includegraphics[width=.55\textwidth]{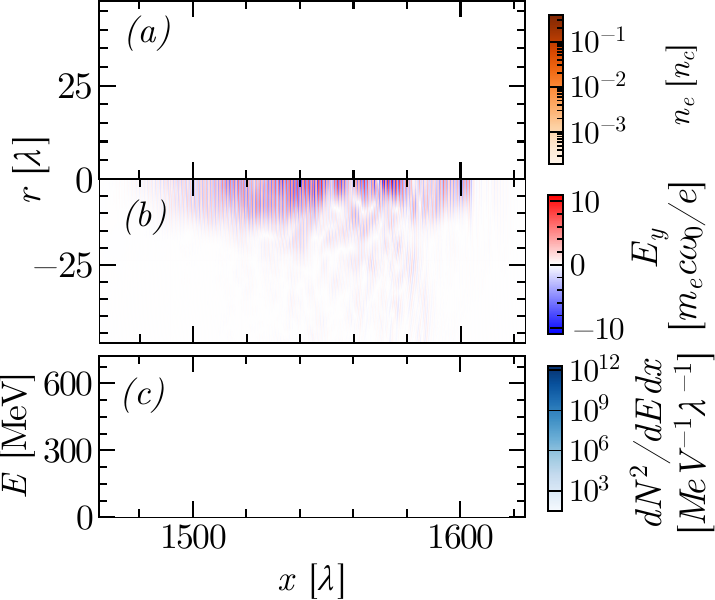}
	\caption{ Illustration of the acceleration conditions for acceleration. The electron density (a), electric component of the laser field (b) and electron energy as a function of longitudinal position $x$ (c)}
	\label{fig:pic_fld}
\end{figure}

The scaling was compared with the quasi-3D PIC simulations using code OSIRIS\cite{fonseca2002,davidson2015}. No guiding structure was assumed in the simulations (the plasma density profile was constant in the transverse direction). The electron density during the interaction, the transverse electric component of the laser field and the energy spectrum as a function of the longitudinal distance are shown in the Fig. \ref{fig:pic_fld}. The energy distribution of electrons is broadband which is a signature of the DLA (as compared to LWFA). Even though the bubble-like structure was created at the laser front, the presence of LWFA is not clearly observed in the $\gamma-x$ plot, as particles get accelerated across the whole laser pulse, most of them "outside the bubble". However, the presence of LWFA during the DLA acceleration is possible and has been observed previously \cite{shaw2017,shaw2018,king2021,miller2023}. It is not a dominant effect in our case. 

\begin{figure}[h]
		\includegraphics[width=.50\textwidth]{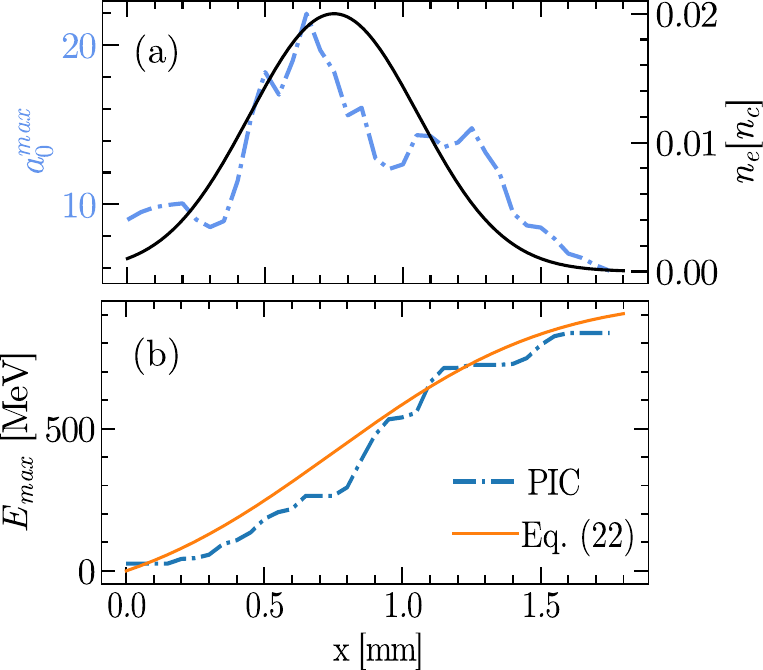}
		\caption{Acceleration in non-ideal laser propagation conditions. (a) Evolution of peak laser amplitude $a_0$ as a function of propagation distance. It has increased during the propagation due to self-focusing. (b) Comparison of the maximum electron energy in the simulation with the Eq. (\ref{eq:sl_gauss}). }
		\label{fig:scaling}
	\end{figure}

The comparison of the scaling law with the results of the PIC simulation can be seen in Fig. \ref{fig:scaling}.  The self-focusing of the laser pulse was significant and the value of $a_0$ increased by a factor of two compared to the vacuum-focused value at a certain point of the pulse propagation. This means that to get an accurate prediction of the electron energy, the value of the maximum $a_0$ as a function of propagation distance should be used in the scaling law instead of the averaged value or the value in a focus. 

In Fig. \ref{fig:scaling} the prediction given by the Eq. (\ref{eq:sl_gauss}) is compared by the evolution of an electron cut-off energy as a function of the propagation distance $x$. The value of $\varepsilon_{cr}$ used for the comparison of the scaling with PIC simulation was $\varepsilon_{cr}=0.2$ which is the same value as observed previously in different studies \cite{babjak2024}. Even though the laser intensity was influenced by the self-focusing during the propagation, the scaling law is in good agreement with the simulation. 

\subsection{Tailoring the density profile to decrease the acceleration distance }\label{sec:pic_shaping}

The cut-off energy of electrons during the acceleration by the DLA can be maximized by guiding the laser pulse over a sufficient distance with the spot size that is matched to the resonant oscillation amplitude for given power\cite{babjak2024}. For example, electron energies up to 2.5 GeV can be achieved with a 1 PW laser, when it is propagated through the plasma of a density $0.1 ~n_c$ and guided with the spot size of $4 ~\rm{\mu m}$. However, it might not be possible to focus a laser pulse to specifically $4 ~\rm{\mu m}$ at some facilities and focusing optics for only a wider spot size might be available. The wider laser pulses achieve the optimal DLA at lower densities, where longer acceleration propagation is needed. To shorten the distance and avoid the situation that the laser pulse would deplete before electrons reach the maximum cut-off energy, structured density targets can be used. At first, laser should interact with the low-density gas plasma matched with the laser power and available spot size for the optimal DLA and when a sufficient number of electrons achieve the resonance, the plasma density can be increased to achieve the acceleration at a shorter distance. Higher plasma density results in the faster energy gain according to the Eq. (\ref{eq:dgdt}). Since we have already demonstrated that the maximum energy of electrons will be given by the properties at the beginning of the interaction, such a tailoring of the density enables the acceleration that reaches the highest possible energies for a given facility at the shortest possible distance.  

We back up our proposed density profile by showing analytically in Section \ref{sec:theory} that the maximum energy that can be achieved at the varying density regime is defined by the local conditions at the moment when electrons start to be resonantly accelerated. In Section \ref{sec:tp_test} we have demonstrated this using a test particle model of electrons accelerated at various density profiles. Even though the profiles were different, electrons achieved the same maximum energy since the plasma density at which they achieved the resonance was the same for all profiles considered.

To verify the advantage of such a density profile we performed PIC simulations for 1 PW and 5 PW lasers interacting with gas jets with accordingly tailored density profiles. The sufficient guiding and control of the spot size was ensured by the preformed quasi-neutral guiding structure. Details of the simulations are summarized in the Table \ref{tab:tab1}. 

\begin{table*}
\caption{\label{tab:tab1} Simulation parameters used for the demonstration of the tailored density profile effects on electron acceleration. }
\begin{ruledtabular}
\begin{tabular}{ccc}
  Laser power  & 1 PW & 5 PW\\
 \hline
  $a_0$& 21.6 & 40 \\
 $W_0$ [$\mu$m] & 10 & 12 \\
 Channel profile & $n_w\left(\frac{r}{R_{ch}}\right)^5 + n_e$   & $n_w\left(\frac{r}{R_{ch}}\right)^5 + n_e$
 \\
 Channel width $R_{ch}$ [$\mu$m] & 25 & 30 \\
 Wall density $n_w$ [$n_c$]  & 2 & 2 \\

 Pulse duration [fs]& 200 &200\\
\end{tabular}
\end{ruledtabular}
\end{table*}

\begin{figure}[h]
	\includegraphics[width=.99\textwidth]{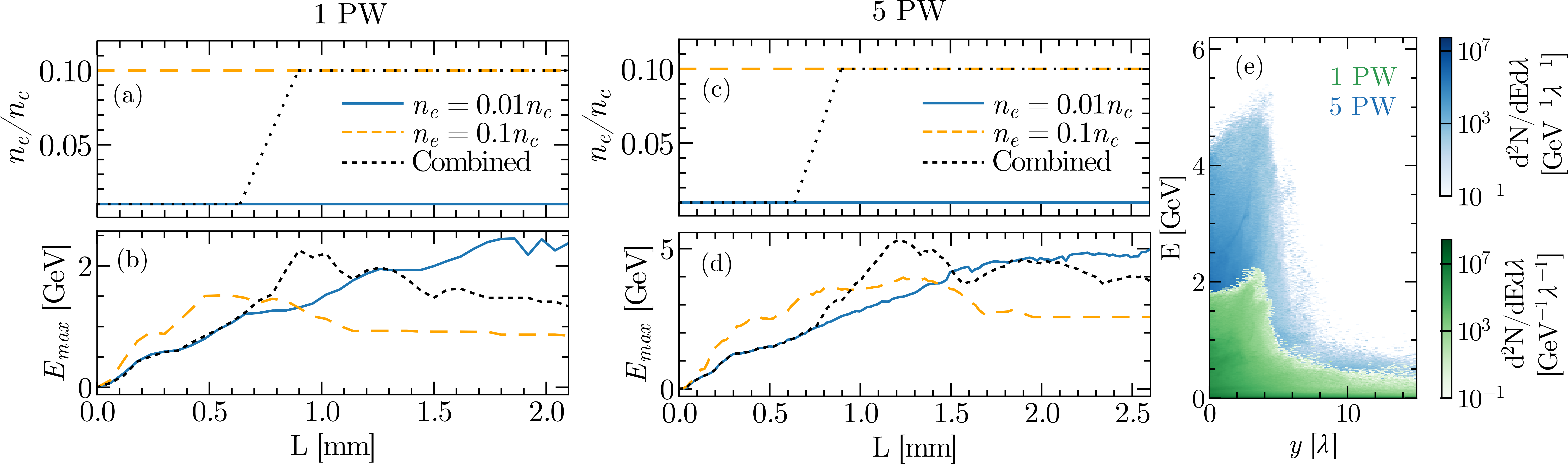}
	\caption{Tailoring the plasma density for reduction of acceleration distance(a) Density profiles in the center of a plasma channel used for the interaction with a 1 PW laser pulse. (b) The evolution of electron energy cut-off during the interaction with 1 PW laser in (a). (c) Density profiles in the center of a plasma channel used for the interaction with a 5 PW laser pulse. (d) The evolution of electron energy cut-off during the interaction with 5 PW laser in (c). (e) The dependence of accelerated electron energies on the transverse distance and the end of the interaction for combined density profiles in (a) and (c).  }
	\label{fig:tailored}
\end{figure}

At first, the laser propagates through the channel with the density $n_e=0.01~n_c$ on the channel axis. The width of the channel and the laser spot size is chosen to undergo the acceleration in the efficient regime which enables to achieve the highest cut-off electron energy possible for the laser with the given power \cite{babjak2024}.  After $\approx 500 \rm{\mu m}$ of propagation through the constant density profile, the background density increases up to $n_e=0.1~n_c$, see Fig. \ref{fig:tailored} (a) and (c). The simulation with a tailored density profile is compared with the simulations of constant densities of $0.1~n_c$ and $0.01~n_c$. 

The acceleration process is qualitatively identical for both examined cases (1 PW and 5 PW). During the interaction with a higher constant density target ($n_e = 0.1~n_c$), the energy gain is fast, but the acceleration stops after the short distance because the maximum energy $\gamma_{max}$ for the given choice of $a_0$ and plasma density was achieved. During the acceleration with a lower density plasma ($n_e = 0.01~n_c$), the acceleration rate is slower which results in a longer acceleration distance but the energy achieved is higher than for the high-density case. The black curve in both graphs of Fig. \ref{fig:tailored} (b) and (d) represents the acceleration at the tailored density profile combining the best properties of both densities. At first, during the low-density interaction, electrons get in the resonance and start to be accelerated with the lower local energy gain. At the moment when the density starts to increase, the energy cut-off starts to increase faster and reaches the maximum energy at a much shorter distance as if it interacted only with the low-density plasma. Note that the maximum energy achieved at the tailored profile agrees with the maximum energy associated with the lower density, which is in agreement with our predictions. This way, the maximum energy was achieved at a much shorter distance which can avoid the problems associated with the need for a laser guiding for long distances. Another advantage of such a profile can be seen in the case of 5 PW, when the energy achieved with a tailored profile was even bigger than the energy achieved by the low constant density target. This can be attributed to the fact, that the acceleration distance needed to achieve the theoretical limit was longer than the depletion length for the laser at a low constant density case.  However, further studies of the laser depletion in the DLA regime are needed to fully describe this effect. 

\subsection{Overcoming the effects of radiation reaction to maximize the electron energy}\label{sec:pic_rr}

\begin{figure}[h]
	\includegraphics[width=.44\textwidth]{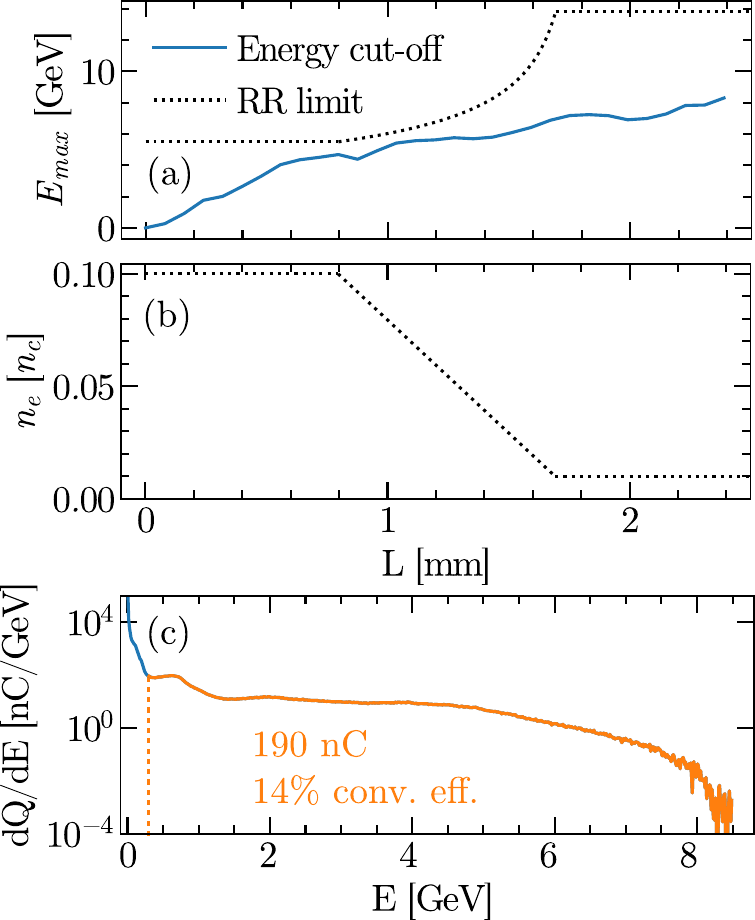}
	\caption{(Acceleration beyond the radiation reaction limit. a) Evolution of the cut-off electron energy during the interaction of a 10 PW and 300 fs long laser pulse with a plasma density profile that aims to achieve the acceleration at the shortest possible distance at the same time as avoiding the damping of the most energetic electrons by the radiation reaction. (b) The density at the center of a preformed guiding plasma channel. (c) The energy distribution function of electrons after 2.4 mm of laser propagation. The total charge of electrons above 300 MeV is 190 nC with 14 \% of laser energy transferred into energetic electrons. }
	\label{fig:rr_ene}
\end{figure}

One of the obstacles in obtaining electron energies near 10 GeV limit is the significant energy loss due to the particle radiation in the field of the background channel. Relatively high gas densities that are relevant for the DLA close to $10^{20}~\rm{cm^{-3}}$ can lead to the creation of focusing channel fields that are the main cause of radiation reaction (RR) losses of accelerated multi-GeV electrons. The limit on the maximum energy that can be achieved by an electron in the radiation reaction dominant regime can be expressed as \cite{jirka2020}

\begin{equation}\label{eq:rr_limit}
    \gamma_{RR} = 1.484\times10^3 \left(\lambda_0 [\rm{\mu m}] \frac{\omega_0^2}{\omega_p^2} \frac{a_0}{\sqrt{2I}} \right).
\end{equation}

The limit given by $\gamma_{RR}$ can prevent electrons from reaching the energies given by the energy limit in the DLA without RR ($\gamma_{max}$ expressed by the Eq. (\ref{eq:gmax})) in a case when $\gamma_{RR}<\gamma_{max}$. One possible workaround is to use lower plasma density and increase the $\gamma_{RR}$ value by decreasing the emitted power due to transverse oscillations. However, long propagation distances at low plasma density can lead to problems with the laser depletion described in the previous section. 

To minimize the problems with laser depletion and to mitigate the role of RR, we propose the following strategy. At first, electrons should start the acceleration at higher densities because of faster energy gain, around the $10^{20}~\rm{cm}^{-3}$. After, when electrons are close to reaching the limit of $\gamma_{RR}$, the plasma density should be decreased to mitigate the impact of RR and increase the value of $\gamma_{RR}$. Density profiles that can be created by two-density gas jets are sufficient as proof of principle, but further optimization and more precise density profile designs can be proposed with the density profile matching the radiation reaction energy limit. 

To test our prediction we run PIC simulation with the density profile shown in Fig. \ref{fig:rr_ene} (b). At first, the laser is focused into the preformed guiding channel with the density $10^{20}~\rm{cm}^{-3}$ and right before electrons reach the RR limit after 0.6 mm, the density on the channel axis was decreased to $10^{19}~\rm{cm}^{-3}$ at $\approx 570$ mm. Laser with the power of 10 PW and duration 300 fs was focused to a spot size 6 microns resulting in peak $a_0=113.33$. The preformed guiding structure had a shape $n_w(r/R_{ch})^5 + n_e$ where $n_e$ is the density at the channel axis, $n_w$ is the density of a channel wall equal to 2 $n_c$ and channel radius $R_{ch}=20\rm{\mu m}$. 

The Fig. \ref{fig:rr_ene} (a) shows the evolution of the energy cut-off and the energy limit given by the Eq. (\ref{eq:rr_limit}). Electron energy exceeded 8 GeV after around 2.5 mm of propagation. By the end, the energy increase was slower than at the beginning not only because of lower density but also because of strong laser absorption. Only around 30\% of the initial laser energy was present in the laser pulse at the end of the interaction. The total accelerated charge of electrons above 300 MeV was 190 nC containing 14 \% of the initial laser pulse energy as shown in Fig. \ref{fig:rr_ene} (c). Tweaking of initial laser focusing, guiding channel and longitudinal density parameters can lead to even higher energies, potentially exceeding 10 GeV limit.

\section{Conclusion}

We have explored the direct laser acceleration mechanism in the case of a slowly varying density profile using analytical theory, test particle simulations in a simplified plane wave geometry and PIC simulations. We have shown that even though the $I$ is no longer a conserved quantity when the plasma density changes, the mean value of the $\gamma_{max} = 2 I^2 \omega_0^2 / \omega_p^2$ is conserved beginning from the moment when electron reaches the resonance. This quantity represents the upper limit for the energy that the most energetic electrons can reach during the acceleration. This means that when an electron reaches the resonance at a certain density, which defines its maximum achievable energy $\gamma_{\rm{max}}$, this energy limit is not changed by slowly varying plasma density profile afterwards. This has been proven to be true by test particle simulations and further verified in Quasi-3D PIC simulations.

Also, the generalized energy scaling law was derived for an arbitrary plasma density profile, with one condition: the profile needs to vary slowly compared to the betatron wavelength. The scaling law can be obtained for an arbitrary profile after substituting the profile into the general formula and integrating it over the interaction path. Specifically, we give explicit solutions for Gaussian and exponentially increasing density profile. This can be directly used to predict maximum electron energies obtained for example in the fast ignition scenario. Also, the scaling law can predict hot electron energies generated in the preplasma of a thin foil used for the ion acceleration via TNSA mechanism. The DLA experiments with expanding plasma plume or by high-density gas jets are also within the scope of the theory. The scaling law for the interaction of a laser pulse with the Gaussian density profile was found to be in good agreement with our PIC simulations. However, one needs to be careful about the validity of the scaling since it assumes constant $a_0$ by neglecting the self-focusing. We have shown that even if self-focusing is present, the scaling can predict a value close to the one observed in PIC simulations. 

Furthermore, we propose a strategy for tailoring the density profile that results in the highest possible electron energy at the shortest acceleration distance. This can be also used to ensure the matched DLA acceleration for various focusing possibilities at a given experimental facility. The strategy consists in starting the acceleration at low density in the matched regime for the efficient DLA to ensure the high cut-off energy. After, the density should be increased to ensure that the energy gain is faster. Following this strategy, we demonstrate using PIC simulation that energies exceeding 5 GeV can be reached after 1 mm of propagation using a 5 PW laser pulse. We also demonstrate the importance of density shaping for a 10 PW laser which will play a crucial role in obtaining electron energies above the 10 GeV limit by the DLA, which would otherwise be impossible due to the radiation loses. 

The results presented in this paper mainly discuss the cut-off energy of the spectrum. However, the generalization to describe the full spectrum is possible if one accounts for electrons that started to be accelerated at later times and electrons located initially closer to the axis. Accounting for all the electrons present in the interaction should allow for assigning the temperature to the hot electron part of the spectrum, which will be addressed in future work. Results can also be applied specifically for particular density profiles associated with regimes such as TNSA, fast ignition, X-ray and gamma-ray generation, or neutron production. Effects such as self-focusing, guiding by preformed plasma channels, superluminal phase velocity, strong laser absorption or instabilities associated with the propagation of relativistic laser pulses through underdense plasma also need to be considered for exact description. We believe that our results provide a significant theoretical basis for better understanding of the above-mentioned regimes of interaction. 
 
\begin{acknowledgments}
The authors acknowledge fruitful discussions with A. Arefiev and L. Willingale. This work was supported by FCT Grants CEECIND/01906/2018, PTDC/FIS-PLA/3800/2021 DOI: 10.54499/PTDC/FIS-PLA/3800/2021 and FCT UI/BD/151560/2021 DOI:10.54499/UI/BD/151560/2021. We acknowledge use of the Marenostrum (Spain) and LUMI (Finland) supercomputers through PRACE/EuroHPC awards. This work was supported by the Ministry of Education, Youth and Sports of the Czech Republic through the e-INFRA CZ (ID:90140).
\end{acknowledgments}

\appendix

\section{Optimal focusing}

It has been previously demonstrated that to use the energy of the laser pulse optimally for direct laser acceleration, it needs to be focused according to the density of the plasma \cite{babjak2024}. During the DLA acceleration, the most energetic electrons are those with high transverse oscillation amplitude. This means that tight focus is not an optimal option since electrons with high oscillation amplitude don't interact with the laser pulse. We also need to keep in mind that the amplitude of the most energetic electrons is proportional to $a_0^{3/4}$. This means, that very wide focusing is not efficient, because it leads to low $a_0$, consequently to low transverse oscillation amplitude $y$ of resonantly accelerated electrons according to Eq. (\ref{eq:ymax}). This results in low energy of accelerated electrons since $\gamma_{max}\sim I^2\sim y^4$.  Here, we aim to show that the same statement holds for the interaction with the varying density profile.

\begin{figure}[h]
		\includegraphics[width=.55\textwidth]{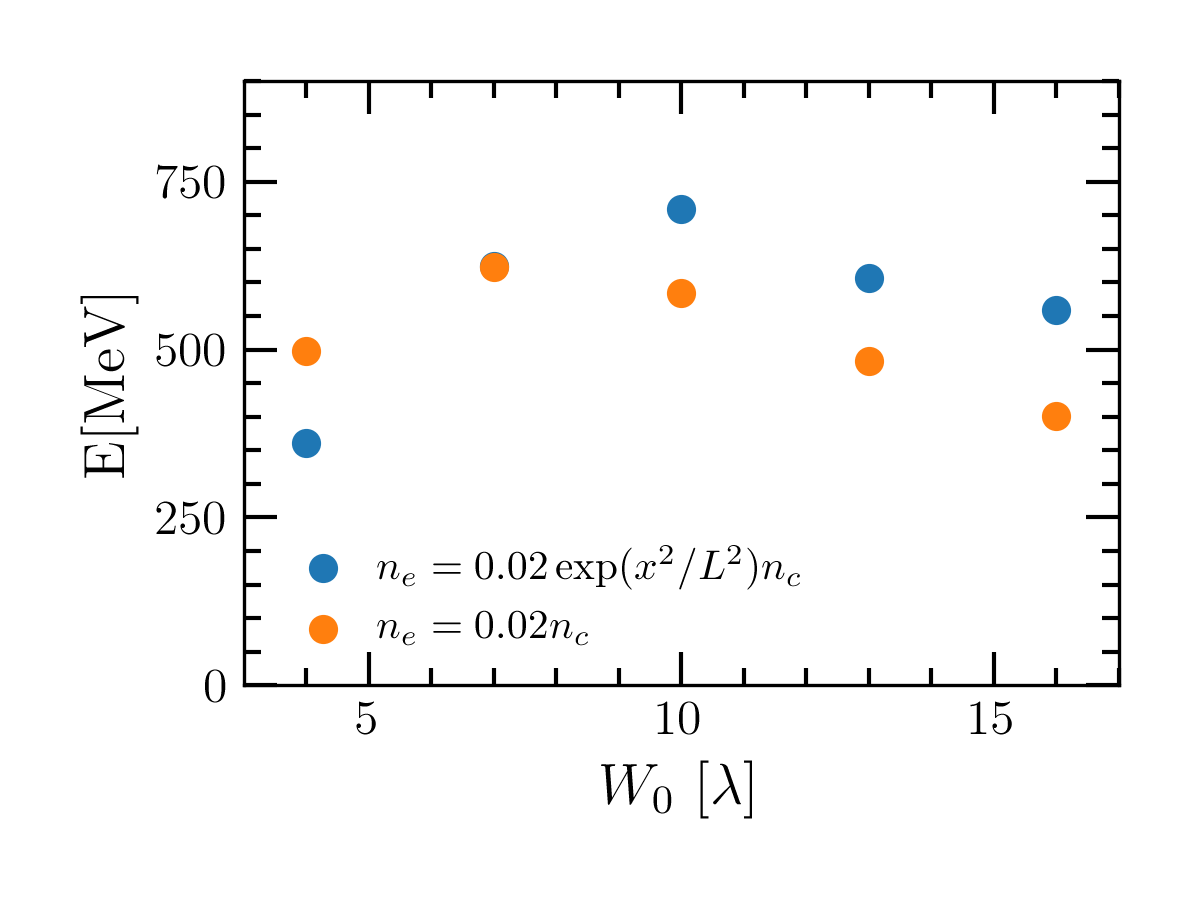}
		\caption{Maximum energies of electrons achieved at simulations for constant power 0.2 PW and varying $w_0$ and $a_0$. The laser pulse interacted with varying and constant density profile for 1.3 mm. }
		\label{fig:w0}
	\end{figure}

To do so we perform a set of Quasi-3D Osiris PIC \cite{fonseca2002,davidson2015} simulations for both constant and a Gaussian density profile with peak plasma density $n_e=0.02n_c$. The laser pulse is chosen to be $\simeq 0.2 \rm{PW}$ and the width $W_0$ was varied while keeping the power constant which resulted in changing the peak $a_0$. Chosen values of $W_0$ were 4, 7, 10, 13 and 16 $\mu \rm{m}$ for $\lambda = 1 \rm{\mu m}$ and resulting $a_0$ were 6.25, 7.7, 10.0, 14.2, 25.0 respectively. The laser pulse propagated for 1.3 mm in both setups. The maximum energies achieved in each simulation are shown in Fig. \ref{fig:w0}. Several conclusions can be made based on the results.

First of all, it is not optimal to aim for the tightest focus possible to achieve the highest $a_0$. The constant density scan showed that the most optimal laser waist was 7 $\rm{\mu m}$ for constant density plasma and 10 $\rm{\mu m}$ for the Gaussian profile. Higher energies were achieved even at the expense of decreasing $a_0$ which is in good agreement with our previous work on optimal focusing\cite{babjak2024}. It's also worth noticing that the optimal width was higher for varying density profile. The reason is, that the Gaussian profile had lower effective density than constant profile. The width of the optimally focused pulse is higher for lower densities according to the theory which agrees with results of our PIC simulations. Furthermore, electron energies were in general higher at the interaction with the varying density profile even though the difference was not too big. This can be assigned to several differences between the two regimes such as different self-focusing, different $\gamma_{\rm{max}}$ of electrons during the acceleration, or injection of electrons with high initial amplitude due to the decrease of the oscillation amplitude due to the increasing plasma density as discussed above.

%\nocite{*}
\bibliography{bibliography}% Produces the bibliography via BibTeX.

\end{document}